\documentclass[iop]{emulateapj}

\usepackage{epsfig}
 \usepackage{psfig}

\def\gtrsim{\mathrel{\hbox{\rlap{\hbox{\lower4pt\hbox{$\sim$}}}\hbox{$>$}}}}
\def\ltsim{\mathrel{\hbox{\rlap{\hbox{\lower4pt\hbox{$\sim$}}}\hbox{$<$}}}}
\newcommand{\noprint}[1]{}

\begin{document}

\title{Characterization of the X-ray light curve of the $\gamma$\,Cas-like 
B1\NoCaseChange{e} star HD\,110432 }

\author{Myron A. Smith}
\affil{Catholic University of America,
        3700 San Martin Dr.,
        Baltimore, MD 21218, USA 
        msmith@stsci.edu}

\author{Raimundo Lopes de Oliveira} 
\affil{Universidade Federal de Sergipe,  
       Departamento de F\'isica, 
        Av. Marechal Rondon s/n, 
        49100-000 S\~ao Crist\'ov\~ao, SE, Brazil \\
}

\and

\author{Christian Motch}   
\affil{Universit\'e de Strasbourg, CNRS UMR 7550, Observatoire Astronomique,
        11 rue de l'Universit\'e, 
        F67000 Strasbourg, France }


\begin{abstract}
HD\,110432 (BZ\,Cru; B1Ve) is the brightest 
member of a small group of ``$\gamma$\,Cas analogs" that emit copious
hard X-ray flux, punctuated by ubiquitous ``flares." To characterize the 
X-ray time history of this star, we made a series of six {\it RXTE}
multi-visit observations in 2010 and an extended observation with the 
{\it XMM-Newton} in 2007.
We analyzed these new light curves along with three older
{\it XMM-Newton} observations from 2002--2003. 
Distributed over five months, the {\it RXTE} observations were designed 
to search for long X-ray modulations over a few months.
These observations indeed suggest the presence of a long cycle 
with P $\approx$ 226\,days and an amplitude of a factor of two.
We also used X-ray light curves constructed from {\it XMM-Newton} observations 
to characterize the lifetimes, strengths, and interflare intervals of 
1615 flare-like events in the light curves. After accounting for 
false positive events, we infer the presence of 955 (2002-2003) and 386
(2007) events we identified as flares.
Similarly, as a control we measured the same attributes 
 for an additional group of 541 events
in {\it XMM-Newton} light curves of $\gamma$\,Cas,
which after a similar correction yielded 517 flares.
We found that the flare properties of HD\,110432 are mostly similar
to our control group. In both cases the distribution of flare strengths
are best fit with log-linear relations.  Both the slopes of these
distributions and the flaring frequencies themselves exhibit
modest fluctuations.
We discovered that some flares in the hard X-ray band of 
HD\,110432 were weak or unobserved in the soft band and vice versa.
The light curves also occasionally show rapid curve drop offs that are
sustained for hours.
We discuss the existence of the long cycle and these flare properties in  
the backdrop of two rival scenarios to produce hard X-rays, a magnetic star-disk
interaction and the accretion of blobs onto a secondary white dwarf. 
\end{abstract}

\clearpage

\section{Introduction: HD110432 among $\gamma$\,Cas analogs}
\label{intrd}

For many years $\gamma$\,Cas (B0.5e\,IV) held the title of ``odd man out"
among X-ray emitting stars because of its peculiar set of X-ray 
characteristics. These include a moderately elevated L$_x$ (by a factor of 
10) compared to X-ray emission of other main sequence B stars and an almost
ubiquitous X-ray flaring.\footnote{We define flares as a local explosion 
of hot plasma that releases the X-rays discussed herein. This definition 
does not necessarily connote a magnetic origin for them as on the Sun. } 
The X-ray spectrum is dominated by emission from an optically thin plasma,
parameterized by $kT$ $\sim$ 13\,keV (Lopes de Oliveira et al. 2010;
``L10"). These characteristics are unlike those in X-ray spectra of other 
massive stars and indicate that an unusual mechanism is at play between 
the Be star and its local environment.

   Recently several members of a new ``$\gamma$\,Cas class" of X-ray
emitters have been discovered with spectral types near B0.5, 
but one, HD\,157832, is classified as late as B1.5e
(Lopes de Oliveira \& Motch 2011).
The first and also brightest of these ``analogs" is the B1e star
HD\,110432 (Smith \& Balona 2006; ``SB"), also known as BZ\,Cru. 
This star is also the object of this study.
The remaining seven members of the class are listed in 
Motch et al. (2007) and Smith et al. (2012; ``S12").

  The prototype of this group, $\gamma$\,Cas, has a high rotation velocity. 
Its spectrum exhibits strong H$\alpha$ line emission strength formed in a 
mass loss disk.   The star in actually a binary system with a 
circular orbit and a P$_{orb}$ = 203.53 $\pm{0.08}$ days (Nemravov\'a et al. 
2012, Smith et al. 2012). Two of the analog members are likely to be blue 
stragglers in Galactic clusters, although none of them is known yet to 
be in a binary system.  These are USNO\,0750-1354972 in NGC\,6649 (50 Myr) 
and HD\,119682 in NGC\,5281 (40 Myr). Our target, HD\,110432, may be a 
member of NGC\,4609 (Feinstein \& Marraco 1979). If so, its age is 60 Myr.
These ages are about twice the age of B0.5-B1 main sequence stars.

  HD\,110432 is situated in the sky at the edge of the Coalsack and
has a revised Hipparcos distance of 373$\pm{43}$\,pc (van Leeuwen 2007).
 Nonetheless, Rachford et al. (2001) found that the ISM column length 
toward the star is an equivalent hydrogen column of 
1$\times$10$^{21}$\,atoms\,cm$^{-2}$.
Lopes de Oliveira et al. (2007; ``L07") and Torrej\'on et al. (2012; 
``T12") have fit the fluxes of {\it XMM-Newton,} {\it Chandra,} and {\it 
Suzaku} high resolution spectra with multi-component, optically thin, thermal 
models. Their models require a larger absorption column(s) toward the X-ray 
sources than the UV-derived ISM column, suggesting that cold matter lies 
in front of them.
The T$_{eff}$ and surface gravity of the star have been 
estimated from a measurement of its optical Balmer jump (22,510\,K, log\,g = 
3.9; Zorec et al. 2005), the strength of the He\,II 1640\,\AA\ line
($\sim$25,000\,K; SB), and a fitting of the
UV-to-IR spectral energy distribution (25,000\,K, log\,g = 3.5;
Codina et al. 1984). Fr\'emat et al. (2005) gave a rotational velocity
$vsin\,i$ = 441$\pm{27}$ km\,s$^{-1}$ for $\gamma$\,Cas. 
SB reported that the velocity of HD\,110432 is at least 90\% of the 
$\gamma$\,Cas value, i.e., $\approx$400 km\,s$^{-1}$.
When we consider their intermediate $sin\,i$ values (Stee et al. 2012, SB),
the $vsin\,i$'s of HD\,110432 and $\gamma$\,Cas suggest that
these stars rotate at close to their critical values.  

  Because of its relative proximity to the Sun, HD\,110432 has been 
the object of more studies than other $\gamma$\,Cas analogs.
The models fitting the X-ray spectra of this star are
similar to the four plasma component model found for $\gamma$\,Cas (S12).
The temperature of the dominant hot component varied from $kT$ = 17\,keV 
in 2002 August to 37\,keV in 2003 January. 
T12 noticed a hard tail out to an energy of 33.5\,keV. 
In their best fitting thermal model they derived a $kT$ of 16--21\,keV. 
Large variations in the hot plasma temperature have not been found in 
$\gamma$\,Cas, despite its more intensive monitoring,
but they are a characteristic of the emission of HD\,110432.
The temperature of HD\,110432's hot component also is higher than 
in $\gamma$\,Cas.  The temperatures of the secondary
components, $k$T $\sim$ 3--8\,keV and 0.2--0.7\,keV by L07 and
Torrej\'on et al. are similar to secondary plasma temperatures found 
for $\gamma$\,Cas (Smith et al. 2004, L10, S12).

 In this paper we add new pieces to the puzzle of X-ray emission 
from this stellar class by utilizing new 
{\it Rossi X-ray Timing Explorer (RXTE)} and {\it XMM-Newton} 
observations of its second best known member, HD\,110432,
to investigate a long term variation as well as flare-like 
events in its X-ray light curve. 

\section{ORIGIN OF THE HARD X-RAY EMISSION}
\label{orgn}

\subsection{Background}

   The current key question in the study of $\gamma$\,Cas stars is: what
mechanism creates its hard X-rays?  L07 pointed out both similarities and 
differences of the X-ray spectra of both $\gamma$\,Cas and HD\,110432 to 
those of active accreting white dwarfs (WDs). We will develop this point 
in $\S$\ref{scwd}.  In addition, the possibility that these stars are blue 
stragglers suggests that binarity is relevant to either ongoing accretion 
onto a degenerate secondary or to past angular momentum transfers to the
Be star from the secondary.

  We begin by outlining those multi-wavelength observations that suggest the
$\gamma$\,Cas's X-rays are formed near the Be star. The best evidence for 
this inference comes from a simultaneous observing campaign in 1996 using 
the {\it Hubble Space Telescope} and the {\it RXTE} and observations with
the {\it International Ultraviolet Explorer} several days later. 
These observations demonstrated that X-ray variations
were correlated with strengths of certain UV lines, such as Fe\,V lines, 
that are expected to form near an early type B star (Smith \& Robinson 
2003; ``SR"). The campaign also demonstrated that rapid dips in a UV 
continuum curve coincided with increases in X-ray flux.  
Both appear to be associated with transparent clouds forced into corotation 
over the surface of the Be star, as do frequent occurrences of migrated
migrating subfeatures ({\it msf}) that are observed in both optical and 
UV line profiles of $\gamma$\,Cas (Smith, Robinson, \& Hatzes 1998, ``SRH").
Also, Smith, Henry, \& Vishniac (2006; ``SHV) found a robust 1.21\,day 
feature in the optical light curve of $\gamma$\,Cas ascribed to rotational 
modulation of a magnetic structure that is rooted well below the atmosphere. 
SRH found no rapid flares in their high-quality UV light curves. 

The correlated rapid variations in the X-ray and UV/optical regimes 
lend further support to the magnetic scenario in $\gamma$\,Cas 
because they suggest that magnetic fields exist on the Be star\footnote{These 
fields would necessarily have complex topologies. Otherwise large-scale 
magnetospheres would be rendered visible through Bp-type modulations 
of the UV resonance line profiles of $\gamma$\,Cas.}
that can interact with inner regions of a decretion (mass loss) disk.
Some of these activities may occur on HD\,110432 as well. However, 
the implementation of multi-wavelength observational campaigns in order to
look for 
them has not been feasible. However, SB did report the existence of {\it msf} 
in profiles of a He\,I line on two nights of a short observing campaign.

 On longer timescales the occurrence for $\gamma$\,Cas
of long optical light cycles of 
length 50-93\,days and amplitudes of a few hundredths of a magnitude has
been demonstrated (Robinson, Smith, \& Henry 2002, ``RSH"). SHV discovered 
ongoing X-ray cycles during 2000--2002,
but with amplitudes of a factor of three. Moreover, these authors found
that X-ray variations over nine days in 2004 that could be 
predicted from contemporaneous optical cycles without 
any {\it ad hoc} parameters other than the ratio of X-ray to optical
amplitudes that RSH had determined for other cycles of this star. 
In 2002 SB found a 0.04 mag. amplitude variation consistent with a 130 day cycle
in the optical light curve of our program star, HD\,110432.\footnote{Optical 
monitoring of HD\,110432 over several years 
by Sarty et al. (2011) did not find a periodicity. However, given their
quoted {\it rms} errors, it is unlikely that they could have discovered
a $\sim$130 day period with the amplitude SB reported.  }  
Herein we also examine new {\it RXTE} data of this object to search for 
X-ray variations on a similar timescale.

 \subsection{X-ray variations in the context of the magnetic scenario}
\label{scnr}

\subsubsection{Long term variations}

A salient attribute of the optical cycles of $\gamma$\,Cas is that their 
amplitudes are smaller in the $B$ than the $V$ filter. This is contrary 
to the ratio $\Delta B/\Delta V$ $\approx$ 1 observed for the star's 
1.21 day period due to the advection of a surface structure. 
The lower amplitude ratios for the long cycles suggest that they are 
excited in an environment cooler than the star's surface temperature, and yet 
it contributes to the Be-disk system's optical brightness. This suggests 
that the optical oscillations arise in the disk.  This inference was 
confirmed recently when Henry \& Smith (2012) observed that the amplitude of
an optical cycle decreased when $\gamma$\,Cas underwent a mass loss episode 
in 2010-2011, and was fully restored after the outburst subsided. 
RSH developed a picture in which the magnetorotational instability (MRI) 
of Balbus \& Hawley (1991, 1998) produces a cyclical disk dynamo. The 
concept envisions a seed magnetic field embedded in the Keplerian disk.
Under these conditions a turbulent dynamo is created that generates cyclical
density changes in the disk. These modulate the 
disk's emission contribution, thereby creating observable light and color 
variations. It is important to stipulate that this is a hypothesis. 
It depends upon the assumption that the dynamo does not consist of many 
independent cells but rather is capable of organizing coherent global disk 
oscillations.

In this dynamo picture the alternating inward and outward matter 
circulation imposes changes to the angular velocity within the disk. 
Even small changes in the angular rate of the disk will impose stresses
on lines of force anchored to the star and interacting with the disk.
Thus, RSH assumed that at certain phases in the cycle the field lines
stretch and break. As these lines reconnect they will relax
and accelerate particles caught within them, electrons most
energetically. Some of these beams will be accelerated toward 
the star where they impact the surface and generate X-ray flares. 
We do not know whether stellar magnetic field lines channel these beams
onto preferred surface areas.  While the 
events leading up to flare creations differ significantly
for the magnetic interaction and WD accretion scenarios, we see the X-ray 
fluxes as the outcome of bombardments from an external source in both of them. 
However, the magnetic disk interaction picture uniquely brings together 
the multiwavelength long-term and rapid correlations, just described, 
with the X-ray flaring, summarized next.

 \subsubsection {\it Constraints from flare observations }
\label{flrconstr}

  The highest time-resolved X-ray light curves of $\gamma$\,Cas 
show that their flares have symmetric profiles in time
and that they have timescales as short as the detection threshold of a few
seconds (Smith, Robinson, \& Corbet 1998; ``SRC"). 
These facts indicate that they must be created in a high density 
environment. The argument for this conclusion is independent of the venue 
and exciting mechanism and  goes as follows: flare emission is a brief event
that manifests the heating of a plasma parcel to an extreme overpressure.
The flare's emission decays either by adiabatic expansion or radiative 
cooling. SRC studied the properties of a large distribution of flares in
$\gamma$\,Cas created at
a high temperature ($kT$ $\gtrsim$ 10\,keV, or over 10$^{8}$\,K) and in an
optically thin medium. They found that expansion effects generally win,
such that even after its Emission Measure decays by one e-folding the now
lower density plasma essentially maintains its temperature. 
For the cases of both the freely expanding and radiatively cooling plasma,
the timescale for the flare decay depends on the reciprocal of the electron 
density, N$_{e}$$^{-1}$. 
See SRC and Wheatley, Mauche \& Mattei (2003; ``WMM") for details.
These arguments suggest an electron density of $\gtrsim$ 10$^{14}$ cm$^{-3}$
for the initial flare parcel.\footnote{Smith et al. (2012) have also
determined average densities of $\gtrsim$10$^{13}$ cm$^{-3}$ from 
density-sensitive line ratios found in high resolution X-ray spectra of
$\gamma$\,Cas. This is consistent with densities derived from flare 
timescales discussed in this subsection.}
Following the plasma's nearly adiabatic expansion, the still hot plasma 
eventually radiates its energy over a much longer timescale. 
SRC associated the hard ``basal" flux component, which has a variability
timescale of 1--3\,ks,  with this plasma residue.

  There are at least two ways of creating flares by means other than
magnetic instabilities occurring on the Be star. One of these is to inject 
high energy electron beams onto its surface from an external source. 
This was the approach offered by Robinson \& Smith (2000; ``RS"), in developing 
the magnetic scenario. A second, WD blob accretion, is described in the 
next section.  The RS models assumed
an almost monoenergetic electron beam ($<$E$>$ $\approx$ 200 keV, i.e., 
v $\sim$ 2000 km\,s$^{-1}$) that originates from an unspecified external 
source and partially penetrates a ``thick target" with a scale height 
of a main sequence B star's atmosphere. 
There the beam heats parcels with radii of $\sim$10$^3$ km to a temperature 
$kT$ $\sim$ 10 keV.  RS's best monoenergetic beam model penetrated to a density 
of 10$^{14}$ cm$^{-3}$ and heated plasma to $kT$$\sim$10\,keV, which are 
close to the observed parameters of $\gamma$\,Cas flares.

 \subsection{X-ray variations in the context of accretion onto a white dwarf}
\label{scwd}

  Could the X-ray properties of the $\gamma$\,Cas variables result
from Be-wind driven accretion onto a WD? One of the difficulties
in addressing this question has been the absence of known Be+WD systems.
Thus, the recent discovery by Sturm et al. (2011) of a candidate
Be (or Oe) + WD system, named XMMU J010147.5-715550, is of some importance.
Located in the Small Magellanic Cloud, this object has a spectral
type of O7-B0 and was detected during a likely nova outburst characterized
by a luminous ``super soft" ($T$ $\sim$10$^{6}$\,K) X-ray emission.
The typical hard X-ray luminosity of $\gamma$\,Cas stars
(L$_{\rm X}\,\sim\,10^{32-33}$erg\,s$^{-1}$) is too weak to be detectable
at the distance of the  Magellanic Clouds.  

However, several of the Galactic magnetic CVs known as ``polars" exhibit 
rapid X-ray variability that can be to some extent compared with those seen 
in $\gamma$\,Cas variables. Perhaps the most similar behavior is that 
of the soft X-ray polar V1309 Ori
(Schwarz et al. 2005) in which
individual flares can be identified. Flare profiles are symmetric with rise 
and decay timescales of the order of 10\,s, hence comparable to those 
seen in HD\,110432. The hard X-ray light curve of AM Her
also displays rapid variations that can be modeled in terms of $1/f$ shot 
noise (Beardmore \& Osborne 1997).  In this case, the modeling favors 
a shot with an exponential decay on a time scale of $\sim$ 70\,s. 
However, the high shot rate implied by the shape of the power 
density spectrum prevents the detection of individual events. 
These flares are usually ascribed to the accretion of individual blobs 
of matter that are pinched laterally and stretched  
by tidal forces as they fall into the accretion column. 
The size and mass of the individual blobs in V1309 Ori
($m_{blob}$\,$\sim$\,2$\times$10$^{18}$ g; Schwarz et al. 2005) are small,
and they penetrate below the photosphere where their shock energy is 
directly thermalized. The hard X-rays emitted there are absorbed by the 
overlying photosphere, rendering them largely unobservable.
In AM Her, blob masses are about two orders of magnitude lower, and as 
such that they cannot penetrate below the photosphere. In contrast to the
large-blob accretors such as V1039 Ori, the shock front stands well above the 
WD surface and its thermalization leads to copious hard X-ray 
emission.  The sensitivity of the peak energies of the X-ray flares to the 
size of the blobs explains the very different characteristic energies 
of different accretors. For an up to date review see Mouchet et al. (2012). 

 The absence of short and stable periodicities in the X-ray light
curves of $\gamma$\,Cas stars indicates that the putative accreting
WD is unlikely to be strongly magnetized. This implies that
direct comparison with the behavior of magnetic CVs is not possible.
In addition, most known accreting WDs channel matter through a
magnetically controlled accretion column to their poles, or when accreting
through the boundary layer of a disc to their equatorial regions.
It is possible that the comparatively low angular momentum of the material
accreted from the outer edge of the decretion disk would lead to spherical
accretion onto the putative WD. However, no
hydrodynamic models have been published yet to test this possibility.

\section{Observations}

\subsection{RXTE light curves}

Although the above description applies to $\gamma$\,Cas, it is not clear
that all the members of this class have identical characteristics because
some of these are difficult to uncover at fainter 
X-ray brightness limits.
To determine which photometric variations are present in
HD\,110432, we mounted a new {\it RXTE} investigation of this object 
by means of Guest Observer Cycle 14 observing time awarded to MAS. 
The program's purpose was to see if a long X-ray cycle exists 
similar to the optical cycle that SB observed in 2002. 

  The {\it RXTE} was a low Earth-orbiting satellite designed to monitor X-ray
sources over many timescales. Its operations were terminated in 2012 January.
It was comprised of two instrumental packages,
the HEXTE for high energies 
and the Proportional Counter Array (PCA) we used for medium energies (2--10 
keV). The PCA consisted of 5 Proportional Counter Units (PCUs) serving as 
detectors, with a total effective area of 6500 cm$^{2}$ at 6\,keV.  
Early in the mission, including for the 1996 observations of $\gamma$\,Cas, 
all five PCUs were active, but in recent years typically only one PCU was used.

\begin{table}[!h]
\tablenum{1}
\begin{center}
\center{\caption{RXTE Observing Log of HD\,110432 (in RJD)}}
\vspace*{.15in}
\begin{tabular}{c|c}
\tableline\tableline
 Obs. 1--3  &  Obs. 4--6 \\
 (start--end)  &  (start--end) \\
\tableline \tableline
55242.8876--55243.2296     &   55340.7623--55341.1501   \\
55268.7822--55269.1283     &   55367.8391--55368.2037   \\
55291.8157--55292.1459     &   55397.6392--55398.0277   \\
\tableline
\end{tabular}
\end{center}
\vspace*{.15in}
\end{table}

Our {\it RXTE} GO observations consisted of 36 satellite orbits 
distributed from 2010 February 15 through July 20.
Table\,1 gives Reduced Julian Dates (RJD) for each epoch.
At each of the six epochs the spacecraft observed our target six orbits 
over an interval of 8-9 hours. Exposures were made using the PSU2 unit 
alone and lasted 1.2--2.7\,ks for each satellite orbit.
Photon events were recorded in all PCU detector layers (except the Xenon layer).
{\it RXTE} ``Standard2" data products were generated by the pipeline processing
system using current calibration and bright-source background files.
Background models were computed using the {\it pcabackest} v3.8, 
which corrects for spike-like artifacts sometimes
found in previous models. The background count rates were 
typically $\sim$50\% of the ``net" (background subtracted) 
rate used for the light curve analysis. Time series of {\it rms} errors 
(mainly caused by photon and electronic noise) were also produced and used 
in our analysis of rapid events in the light curves.
``Standard2" data thus include a time series (light curve) binned to 16\,s 
for each satellite visit.  

\begin{figure}[h!]
\centerline{
\includegraphics[scale=.35,angle=90]
{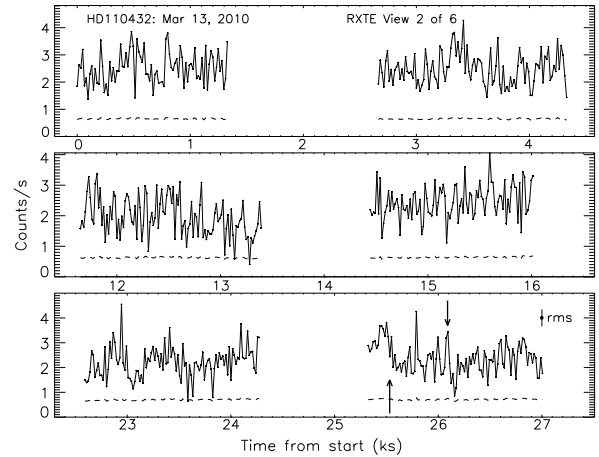}}
\caption{
Light curves from 6 {\it RXTE} exposures of HD\,110432
on 2010 March 13 (RJD\,55268). These data represent the second
epoch shown in  Fig.\,\ref{lclong}.
Dashed lines are the RXTE pipeline-generated errors.
Data gaps are due to Earth passage near the target
sightline. The largest fluctuations are blends or ``aggregates" of X-ray
flares. One such feature is shown by the downward pointing arrow at 26.1\,ks.
An upward arrow at 25.55\,ks shows the sudden drop in count rate near the
beginning of the last exposure.
\label{xtelc}
}
\end{figure}

Figure\,\ref{xtelc} exhibits an example of light curves for the epoch
2010 March 13.  In this rendition most of the jagged peaks are groups 
of unresolved flares we call aggregates. These have a typical timescale 
of a few minutes.
We will discuss other features of this light curve in $\S$\ref{reslts}.
In general, our analysis indicated that {\it RXTE} light curves have 
a sufficient data quality for describing long and intermediate timescale
variations but not for studying individual flares.

\subsection{XMM-Newton light curves}
\label{xmmcntrl}

  To study intermediate and rapid variations of HD\,110432 we used four
archival {\it XMM-Newton} observations as our primary data source.  
Two were carried out in 2002 and one in 2003.  
The fourth observation, originally granted to CM as PI, was executed in 2007. 
Four shorter observations of $\gamma$\,Cas were conducted for our group in 2010.
We note that we chose arbitrarily the first of the observations 
of HD\,110432 as a reference dataset our our flare study of this star. 
We used the first three of the four $\gamma$\,Cas datasets as a ``control" 
for the HD\,110432 observations. (We chose these three because they 
cover a total timespan near 50\,ks, i.e., about the
duration of the 2002--2003 HD\,110432 datasets.)
All these XMM-{\it Newton} observations are summarized in Table\,2. 

 The spectra of $\gamma$ Cas obtained from Table\,2 have been discussed
by Smith et al. 2012. This paper on timing analysis of HD\,110432 
supplements that work by using flare events in the $\gamma$\,Cas as a
reference and comparison.

\begin{table}[!t]
\tablenum{2}
\begin{center}
\center{\caption{XMM-{\it Newton} Observing Log}}
\vspace*{.15in}
\begin{tabular}{cccc}
\tableline\tableline
 Target & Obs. ID & Start time & T$_{exp}$ (ks)  \\
\tableline \tableline
HD 110432 & 0109480101 & 2002-07-03 14:59:24.0 & 49.4\\
& 0109480201 & 2002-08-26 21:03:06.0 & 44.8\\
& 0109480401 & 2003-01-21 00:15:30.0 & 44.4\\
& 0504730101 & 2007-09-04 14:55:16.0 & 80.3\\

$\gamma$ Cas$^*$ & 0651670201 & 2010-07-07 07:08:50.0 & 17.5\\
                 & 0651670301 & 2010-07-24 03:31:41.0 & 15.7\\
                 & 0651670401 & 2010-08-02 13:49:58.0 & 17.5 \\
\tableline
\end{tabular}
\end{center}
\begin{list}{}
\item $^{*}$Used as a control for flare measurements.
\end{list}
\vspace*{0.08in}
\end{table}

The archival 2002-2003 XMM-{\it Newton} observations of HD 110432 
allowed the use of {\it EPIC} cameras, but not of the RGS cameras 
because of the off-axis position of the star in the camera's field of view. 
A timing resolution of 200 ms was set as part of the
{\it extended full window} mode used for the pn camera in all observations.
The observations were described in detail by L07, where spectral and 
some timing analyses were done.
This paper supplements the timing analysis in that work.

 The  2007 observation of HD\,110432 was split into two exposures
(80.3\,ks + 9.1\,ks).  However, only ObsID 0504730101 (80.3\,ks) had 
a low enough background to be suitable for analysis. 
This exposure was made with the EPIC cameras
operating in {\it small window} mode through the pn camera 
during 2007 September 04-05. This resulted in a time resolution of 
5.7\,ms for this camera.

Although RGS camera data were also obtained for HD 110432 in 2007 and 
$\gamma$ Cas in 2010, they were not suited to the present study because of 
their low efficiencies and the nonnegligible (4.8\,s) total CCD readout times 
(Pollock 2011). The MOS camera data were not used because of the low 
signal-to-noise values they suffered for the high time resolution we needed 
in this work.  Thus, this paper is based only on EPIC-pn data. 
Also, we note for all our {\it XMM-Newton} observations
that the backgrounds are negligible for timing analyses. 
For example, even in the case of our highest background observation
of HD\,110432, ObsID 0109480201,
the mean count rate of the background is 0.02 counts per second over a 
chosen energy range of 0.2--12\,keV, 
whereas the mean rate of the net spectrum is a few counts per second. 
To demonstrate this point, we will depict the background rates in 
two examples of light curves to follow. 

 All data were reprocessed using the {\it epproc} task running in SAS
v10.0.0, applying calibration files made available in 2011 January. 
The light curves were extracted and background subtracted with the
{\it epiclccorr} SAS task. Flux errors were generated by the processing 
pipeline and utilized in our flare analyses described below.

\section{Results: Characteristic Variations}
\label{reslts}

\subsection{Search for a long cycle with {\it RXTE}}
\label{srch}

\begin{figure}
\centerline{
\includegraphics[scale=.35,angle=90]
{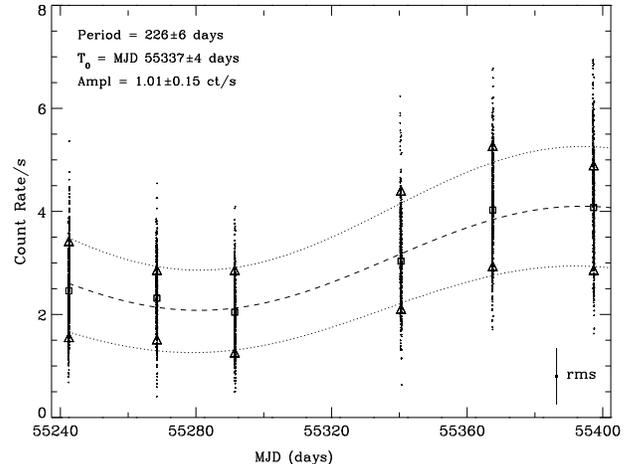}}
\caption{ 
Count rates for the 6 {\it RXTE} visits to HD\,110432 in 2010.
Altogether, 3810 16\,s data points are indicated. 
The typical error bar refers to the mean {\it rms} for one 16\,s integration. 
The dashed smoothed curve is a sinusoid passed through the observations. 
Dotted lines are sinusoids passing through the lower and upper 10 
percentiles (triangles) of the distribution of data. The dashed line is
the sinusoidal solution through the median data points (squares) of six 
{\it RXTE} observations.  The periods and zeropoints for these
latter curves were used to determine the indicated errors. 
\label{lclong}
}
\vspace*{.08in}
\end{figure}

   Figure\,\ref{lclong}
exhibits the light curve for the entire set of {\it RXTE} observations over
a timespan of 155 days. A sinuous trend is obvious in the time series.
We computed median averages of the 36 orbital data sequences and fit a least
squares sinusoid through the full light curve. The fitted period is 226 days. 
We also determined 10\%-tile and 90\%-tile count rates for each orbit
and determined least squares sinusoids through them. 
The two sinusoids gave periods six days different from the median count
rate result of 226 days, so we have adopted a standard deviation 
of ${\pm 6}$ days for our period.  The sinusoid has a mean value 
of 3.09${\pm 0.05}$ ct\,s$^{-1}$ and a semi-amplitude of 1.01${\pm 0.08}$ 
ct\,s$^{-1}$, giving a flux ratio L$_{x_{max}}$/L$_{x_{min}}$ of 2.0.

  One cannot really claim that the variations shown in 
Fig.\,\ref{lclong} represent a complete proper period because only 
$\approx$70\% of it has been observed.
Even so, we can compare the attributes of this X-ray variation with those
of the optical variation SB found and also with the patterns in 
long X-ray and optical cycles of $\gamma$\,Cas to see whether they are 
mutually consistent with the long cycle description.

First, the 226 day X-ray cycle length HD\,110432 is longer than 
the 130 day optical cycle found by SB in the same star
by a factor of 1.7.  Coincidentally or not, this is nearly the same ratio 
as the range of optical cycle lengths for $\gamma$\,Cas, 1.8 (91 days 
over 50 days; SHV, Henry \& Smith 2012).
Then the similar ratios of the lengths of these cycles 
in both optical and X-ray regimes suggest that the cycles
in the two star-disk systems could be produced by the same physical 
mechanism.  Second, if we take for our {\it RXTE} light curve the median
point-to-point count difference as a measure of ``noise"  
and the ``signal" as the median count rate for each of the {\it RXTE} 
epochs, we can compute a pseudo-``signal-to-noise ratio" (SNR) 
for the X-ray cycle
of 20. For $\gamma$\,Cas (1999-2000 seasons) a corresponding pseudo-SNR has
a similar value of 15 (RSH). Third, if one compares the amplitude of the 
X-ray sinusoid for HD\,110432 to the amplitude of the optical cycle, their
ratio (L$_{x_{max}}$/L$_{x_{min}}$)/$\triangle V$ is 67. This is close to
the ratio of 80 RSH found for the simultaneous optical and
X-ray cycles of $\gamma$\,Cas
in 1999--2000.
Each of these arguments is consistent with the long sinuous variation of 
HD\,110432 being part of a sinusoidal long cycle whose attributes are 
similar to those of $\gamma$\,Cas, except that the periods are longer.

\subsection{X-ray luminosity estimate from XMM-Newton}

Assuming a distance of 373 pc, we derive L$_{x}$ = 7$\times$10$^{32}$ 
erg\,s$^{-1}$ (taken over 0.2--12\,keV) from the 2007 {\it XMM-Newton} data.  
This falls within the range of 3.9--8.0$\times$ 10$^{32}$ erg\,s$^{-1}$ 
that L07, Torrej\'on \& Orr (2001; {\it BeppoSAX}), and Tuohy et al. 
(1988; {\it HEAO-1}) found for similar energy ranges. 
The range of L$_x$ derived from the six epochal {\it RXTE} observations 
shown in Fig.\,\ref{lclong} was 3.5--8.8$\times$10$^{32}$ erg\,s$^{-1}$.
RSH used the {\it RXTE} to determine an X-ray luminosity range for 
$\gamma$\,Cas of L$_x$ = 3.1--8.6$\times$10$^{32}$ erg\,s$^{-1}$, 
again using a revised Hipparcos distance this time of 168 pc. 
This puts the mean X-ray luminosities of the two objects within the
same range.

\subsection{Power Density Spectra of XMM-Newton Observations}

 The power density spectrum (PDS) is traditionally introduced to
help evaluate X-ray light curves that exhibit flux variations 
on a variety timescales, particularly if there is reason to believe that
more than a single mechanism is at work to cause them. Thus, our discussion 
of rapid variations starts with a comparison of the slope of the high 
frequency signal with $1/f$ ``red noise." The behavior of the PDS can give 
hints as to additional phenomena, e.g., flare decay timescales, the 
evolution of large scale features, and/or rotational modulation.

\begin{figure}
\centerline{
\includegraphics[scale=.35,angle=90]
{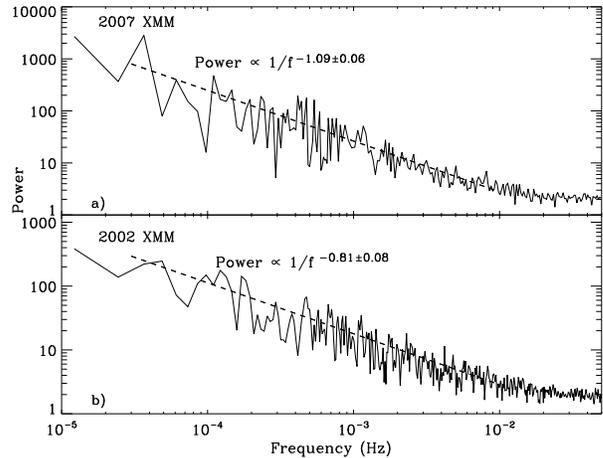}}
\caption{
The power density spectra (PDS) of light curves of HD\,110432 observed by the 
{\it XMM-Newton EPIC} camera during 2007 (panel a) and three combined
visits during 2002--2003 (panel b).  Power at frequencies below the point 
where white noise dominates at $\sim$0.01 Hz is dominated by intermediate 
timescale undulations and not by X-ray flares. The mean slopes 
fitting the PDS are significantly different during these two epochs.
\label{pwrsp}
}
\vspace*{.08in}
\end{figure}

 Toward this end we binned the time series data to 5\,s cadences
over an energy range 0.6--12\,keV from photon list products processed from 
our {\it XMM-Newton} observations of HD\,110432 to create the PDS.
This function, exhibited in Figure\,\ref{pwrsp}a for the 2007 dataset, 
indeed shows an almost $1/f$ dependence at intermediate signal frequencies.
To see if a red noise-like logarithmic slope of -1.09${\pm 0.06}$ 
can be found in other HD\,110432 light curves,
we computed power spectra for the three 2002-2003 observations 
and summed them. The resulting spectrum is shown in Fig.\,\ref{pwrsp}b. 
Its slope is only -0.81${\pm 0.08}$. Thus it is probably coincidental 
that the slope for 2007 was nearly the red noise value of -1.0.
We remark that the -0.8 slope is the shallowest yet found in
power spectra for either HD\,110432 or $\gamma$\,Cas.  (For $\gamma$\,Cas
light curves the spectra must often be fit with two segmented lines; e.g., 
Smith et al. 2012.) Another feature in these spectra is a small dip for
frequencies just below 1$\times$10$^{-4}$ Hz, 
corresponding to $\gtrsim$3 hour variations. 
This feature is marginally significant and begs for confirmation. 
The PDS from the new {\it Chandra} light curve published by T12 
shows a similar dip in power just below 1$\times$10$^{-4}$ Hz, 
though the feature is less clear in their {\it Suzaku} PDS. 

\subsection{Flares in the HD\,110432 light curves}

\subsubsection{Flare measurement procedure}
\label{flrmeas}

  Because the quality (SNR) of the {\it XMM-Newton} datasets was  
essentially identical for
the 2002--2003 and 2007 epochs, we were able to use them all to compile
unweighted statistics of flaring events.  We determined that
5\,s binning is about the practical minimum needed to evaluate flare
profiles, given the {\it EPIC-pn} count rate of this source. It is
also about the same cadence (4\,s) that RSC used in their
analysis of $\gamma$\,Cas flares. We note in passing that the flare
rates probably
would increase if we could use even more sensitive instruments to detect 
{events with shorter timescales than 4-5 s.

  Our strategy was to evaluate the following flare quantities: 
integrated count (similar to a spectral equivalent width), 
lifetimes, and interflare intervals.
To perform the analysis we wrote an interactive computer program 
{\it flarecount} in the {\it Interactive Data Language} to compute these 
quantities via a semi-automated, menu-driven process and a second program 
to edit the history of trial runs on individual flare candidates.  
{\it Flarecount} was executed interactively on events in consecutive 1\,ks  
(200 5\,s bin segments shown on an computer screen surveyed 
along the light curve. 
Our display showed typically 10-20 data bins beyond 
either edge of the screen window to 
define a  ``background" if flare candidates occurred at the edges of the 1\,ks 
field.  In analogy to determining a spectroscopic equivalent width relative to
a spectral continuum, we computed a smoothed undulating baseline 
(referred to as the basal flux in $\S$\ref{flrconstr}). 
This baseline was determined as a $X$th percentile average 
over a contiguous set of data bins we call a ``superbin." 
We then interpolated 
between these artificial points to define a reference baseline. 
A value $X$ = 45\% was chosen and frozen for all measurements. 
This selection is discussed in $\S$\ref{bsln}.
Typically we found that a superbin length of twenty 5\,s bins fits the 
underlying baseline curve well. The superbin value varies along a light
curve because of changes in the ubiquitous low-amplitude 
undulations on a timescale of 5 minutes to about an hour. 
Depending on the timescale of this variation, a superbin ranged from
10 to 40 bins. 

Our program also computed error ({\it rms}) fluctuations from the 
count rates
and used them if they were larger than the smoothed value of the {\it rms} 
vector supplied by the pipeline.  
We used this {\it rms} amplitude as a metric to define positive 
fluctuations as ``flares." An event was considered a 
flare if the fluctuation remained above one {\it rms} above the baseline
for at least two consecutive time bins. We defined the flare strength as
the integral of the count rates in excess of the baseline over time.
The flare duration was then defined as the time over which the 
profile remained one {\it rms} above the baseline level. 
The shortest-lived flares we could resolve in our 5\,s time binned data
had durations of 12.5\,s (FWHM).  On average we found that the FWHM we 
measured is $\approx$80\% of the full duration of a flare.
We saved the input local superbin value as well as the computed 
flare parameters, so we could repeat the measurements if needed.


\begin{figure}[h!]
\centerline{
\includegraphics[scale=.35,angle=90]
{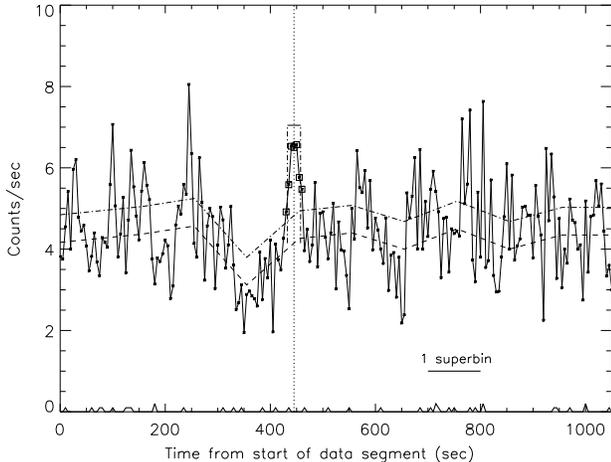}}
\caption{
Example of an 
extraction of a flare at time 445\,s (dotted line) in the 2002 Obs. 1
X-ray light curve. 
The binned background count rate is shown at the bottom of the plot.
The integrated count and FWHM duration of the flare 
indicated are 14.5 counts (1.5$\times$10$^{33}$ ergs s$^{-1}$) and 32.5\,s. 
The lower dashed line are the 45th percentile count rates;
the upper one represents levels 1\,{\it rms} above them. The computed flare 
strength in units of detector counts is denoted by a rectangle. This rectangle
represents the integrated flare strength.
Square symbols are observed fluxes enclosed within the rectangle width. 
\label{flrprf}
}
\end{figure}


  Because the selection of superbin lengths depends on unpredictable
changes in the background variation timescale,
in principle a personal bias can be introduced to the measurements.
To minimize this bias,  we used the reference (2002 Obs.\,1) light curve, 
first measuring its flare events and then remeasuring them after measuring
the Obs.\,2 light curve. The first and second measures produced 
the same number of flares within 5\%.
To minimize systematics further, we rotated the order of time
segments measured among the 2002-2003, 2007 datasets of HD\,110432 and
the (control) dataset of $\gamma$\,Cas.

\subsubsection{ Checks on errors in our procedure}
\label{bsln}

  After establishing repeatable procedures for our
measurements, we investigated how systematic and random errors
influence our results.  To do this we repeated our initial 
set of measurements for our reference light curve. This was
easy to do using the saved superbin lengths 
from our initial measurements and repeating the measurements using 
input values of $X$ = 40\% and 50\%
(50\% means no bias from flare or flare aggregate signals)
rather than 45\%. A change in $X$ causes predictable changes in 
the integrated flare counts. The alternative $X$ values caused changes 
to the measured flare strengths averaging +1.1 and -1.0 counts for 
the overall flare numbers and 2--3 counts for $\ltsim$10\% of 
the largest ones. 
The flare strength distributions for all three assumed values of
$X$ are exhibited in Figure\,\ref{flrbsln}. In this figure the 40\% and 50\%
histograms were shifted by -1 and +1 counts, respectively. The overlap
in these distributions is surprisingly good and indicates that flare
properties other than their integrated counts are almost the same if we
adopt a different $X$ level. From our experience going through this exercise 
manually we estimate that these values of $X$ are upper bounds. We estimate 
an error of the mean $X$ of ${\pm 2-3}$\% of the flux level, i.e., it lies
in the range 42--48\%. The total of integrated counts of our flare
events is also consistent with this baseline.

  We also considered the effect of the integrated count
threshold on the number of detected flares. 
Using the above measurement criteria, we adopted an empirical 
threshold of 3.7 counts. In fact, the numbers of flares are 
insensitive to this threshold:  to decrease the number of flares 
by 10\% would require an increase of 40\% in the threshold count rate. 
Then as long as we maintained the same threshold in our measurements, 
the errors in our flare counts remained comparatively small. 
Erring on the conservative side, we assign an additional 5\% error
in our flare totals from uncertainties in the threshold count value. 
We add this in quadrature with another (likely generous) 5\% error for 
potential personal errors to estimate an final error $\sigma$ = 8\% for the 
measured flare numbers in the {\it XMM-Newton} light curves.

\begin{figure}
\centerline{
\includegraphics[scale=.35,angle=90]
{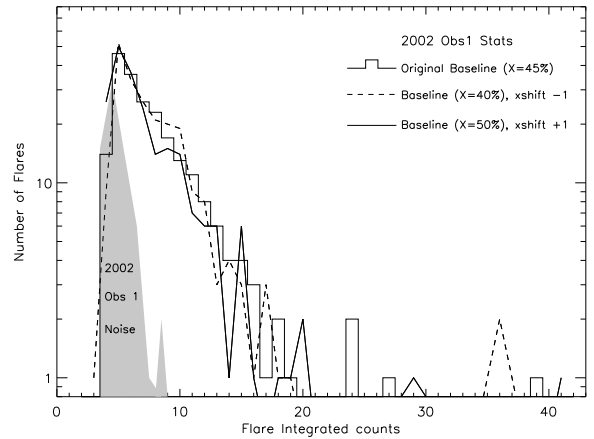}}
\caption{
The distributions of measured flare energies (in {\it XMM-Newton}/EPIC 
counts) for the 2002 Obs. 1 light curve, given three assumed
local baseline energies: $X$= 45\% (histogram, and 
the adopted value), $X$=40\% (dashed line), which is shifted horizontally 
by -1 counts in this diagram, and $X$=50\% (solid line), shifted by +1 count. 
The gray area represents ``false positive" events measured from the 
negative count rate variations in this light curve.
\label{flrbsln}
\vspace*{0.05in}
}
\end{figure}

As for the importance of random errors masquerading as astrophysical flare 
events, a key test of the reliability of our measurements was to determine the 
importance of ``false positives" due to photon event fluctuations 
or possible instrumental or environmental variations.
We quantified this test by repeating the measurements for our reference 2002 
Obs.\,1 light curve yet again and assuming Gaussian distributed errors. This 
time we used the identical superbin lengths and nearly all other 
parameters
from our first measurements,
except that we measured {\it negative} fluctuations instead.
(The one difference was that we adopted an unbiased value $X$ = 50\% for
our baseline.) The advantage of this technique was that all possible
signal properties (especially the sinuosity of the baseline) are 
treated exactly the same way as any possible personal equation 
introduced to the measurements.  Unsurprisingly, 
we found many occurrences of 2 or even 3 consecutive 
negative fluctuations
that were lower than the local mean rate by about 1 {\it rms}, but 
far fewer triggered a false positive that give an integrated count rate 
algebraically less than -3.7 counts.  In all we measured 71 negative 
detections and tabulated their (negative) count rates and durations. 
They are exhibited as a gray area in Fig.\,\ref{flrbsln}. 
These should be subtracted against the total measured flare numbers, 
giving a false positive rate of 17 ${\pm 3}$\%.  
The result shows that the false detections nicely track the lower envelope of 
so-called weak flares of the measured distribution up to about 5 counts. 
Essentially all the low envelope positive-fluctuation events in this figure 
should therefore be regarded as false. The false events fall off quickly toward 
higher energies, and for 6 counts and above, nearly all the measured events
can be regarded as astrophysical. 


\subsubsection{Flare statistics from XMM-Newton light curves}
\label{flrstat}

We also searched for flares in the 2007 time series of HD\,110432.
However, we limited ourselves to the first 50\,ks of the 2007 data to 
keep the flare statistics errors about the same as for the 2002-2003
datasets.  The 2007 data measurements identified 465 flares. 
Correcting for false positives, this corresponds to
an estimated rate of 0.0077 flares s$^{-1}$ (386 flares).  
Our three light curves in 2002--2003
had exposure lengths of  44 to 49\,ks, and we found totals of 418,
397, and 335 flares in the three time sequences, respectively.
Accounting for a 17\% false positives rate, the inferred number of flares 
detected for observations in 2002--2003, 1150, should actually be 955. 
Then the mean flare rate for all three datasets is 
0.0071$\pm{0.0006}$ flares s$^{-1}$, 
while the means of the three of them are consistent 
with the overall mean. These are slightly lower than or comparable to the 
2007 epoch rate. Including the 2007 data, the false positives-corrected mean 
rate is 0.0072 flares s$^{-1}$ } 

 The three {\it XMM-Newton} light curves of $\gamma$\,Cas we used
as controls ($\S$\ref{xmmcntrl}) each had timespans of 15.7-17.5\,ks and
thus their total length is close to the durations of 
each of the 2002--2003 HD\,110432 light curves. 
We measured a mean event rate of 0.012 s$^{-1}$ for $\gamma$\,Cas,
corresponding to a total of 541 measured flares. For the superior 
$\gamma$\,Cas data the false positives rate was 4.5\%.
We therefore subtracted off 24 presumed false positives from the 541 
event total, yielding an inferred net of 517 flares. 
However, the signal-to-noise ratio for these data was a factor of three
higher because $\gamma$\,Cas's higher X-ray flux. 
Therefore, we degraded the SNR of these three light curves with artificial
noise in order to match the mean SNR of the HD\,110432 sequences.\footnote{
The ratio of the mean count rates for $\gamma$\,Cas to HD\,110432, and
averaged over the three {\it EPIC} cameras is a factor of 11.} 
These results were employed to compensate for the differences in X-ray flux
and time binning in the datasets for the two stars. The SNR-degraded light
curves were again then run through the {\it flarecount} program to identify 
a new total of only 450 flares, giving a frequency of 0.0089 flares 
s$^{-1}$.  Again applying the 17\% false positives rate determined from
HD\,110432 to this degraded data, the rate for the $\gamma$\,Cas data 
decreases to 0.0074 flares s$^{-1}$. Within the errors this is the same rate
as the mean determined for HD\,110432. 
Similarly, we degraded the $\gamma$\,Cas 
{\it RXTE} light curve signal for 1996 March (SRC), 
rebinned to 5\,s, and ran {\it flarecount} on it to find 201 flares 
over the same interval. This gives nearly the same mean (uncorrected) frequency
of 0.0090 flares s$^{-1}$ as the 0.0089 flares s$^{-1}$ figure just noted.
Thus, when corrected for the difference in signal properties, the flare rates
of the various light curves of HD\,110432 are about the same. 

\begin{figure}
\centerline{
\includegraphics[scale=.35,angle=90]
{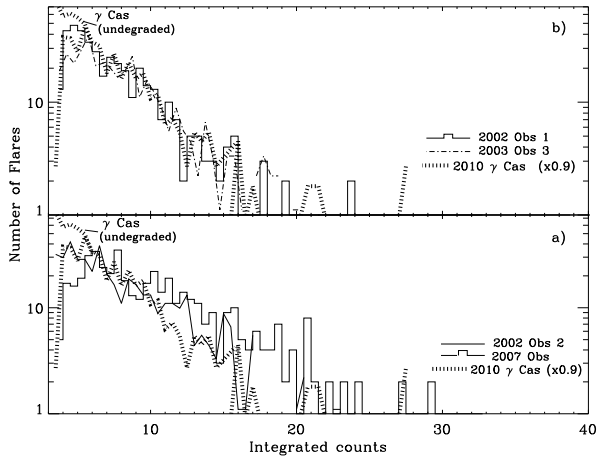}}
\caption{
  The distribution of measured flare strengths (in {\it XMM-Newton}/EPIC 
counts).  All curves in this and following figures are scaled to the number of 
flares occurring within the 2002 Obs.\,1 exposure length (49.4\,ks). 
Thick dashed lines represent two distributions from 2010 light curves of 
our control, $\gamma$\,Cas: the labeled ``undegraded" curve denotes
the numbers measured from the original data binned to 5\,s, while the
(unlabeled) SNR-degraded distribution turns down at low energies.  
{\it Panel a:} the HD\,110432 distributions for 2002 Obs.\,1 and 
2003 Obs.\,3.
{\it Panel b:} the HD\,110432 distributions for 2007 and 2002 Obs.\,2. 
\label{flrew}
\vspace*{0.15in}
}
\end{figure}

  An important characteristic of the flares is the distribution of their
energies, which differs from a power law.
These distributions are exhibited logarithmically in Figure\,\ref{flrew}; 
flare strength units are {\it XMM-Newton/EPIC} counts$^{-1}$.  
Both panels show that the incidence of flare strength (integral over 
the flare-enhanced count rate) is a monotonically decreasing function.
The turnover that exists at low strengths is an artifact of the
limited photon statistics. Using the $\gamma$\,Cas data again as a control, 
this turnover (occurring at $\approx$5 counts) 
is shown by the upper thick dashed lines 
representing the flare distribution taken from the undegraded 
(as observed) light curves of $\gamma$\,Cas, and as the lower
dashed lines from the SNR-degraded light curves. 
We have computed slopes for the HD\,110432 flare  
distributions shown in Fig.\,\ref{flrew} over the abcissa range 6--19 
counts. The slope in the log-linear space is -0.065 ${\pm 0.011}$ for the 
2007 dataset and -0.011, -0.012, 
and -0.015 (${\pm 0.002}$ in each case) 
for the 2002--2003 flares, respectively. 
These rates are only a little affected by false positive events, most
of which occur below the 6 count limit we used to measure the slopes. 
In any case the relative slopes are almost completely unaffected.
These values suggest that at a significance of over 4 sigmas the 2007
flares follow a flatter distribution than the 2002--2003 ones. 
The 2003 flares have a marginally steeper distribution in turn than the 2002 
ones. This suggests that the flare distributions can change slightly on 
timescales of about a year. RS found a similar range of slopes in flare 
distributions over a similar timescale.

\begin{figure}[h!]
\centerline{
\includegraphics[scale=.35,angle=90]
{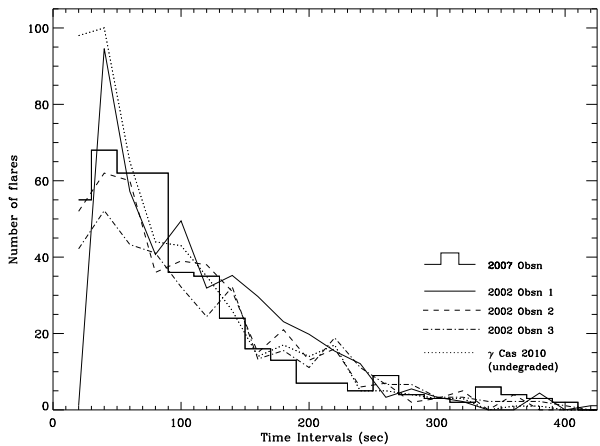}}
\caption{
The distribution of interflare intervals extracted from {\it XMM-Newton}
light curves of HD\,110432 and (undegraded) $\gamma$\,Cas. Note the
turn over at $\sim$40\,s in the HD\,110432 distributions.
\label{flrint} 
}
\end{figure}

  For HD\,110432 the distributions of interflare intervals in 
Figure\,\ref{flrint} are consistent with an exponential distribution for
times $\gtrsim$40\,s, and thus they appear to occur at random times. 
The turnover at this time is 
partially an artifact caused by the limited SNR of the light curves. 
This is to say, the photon noise sometimes encourages us to measure 
closely spaced flares as single unresolved ones.
(We omitted the strongest flares in our measurements of the logarithmic 
slopes of the flare strength distributions in Fig.\,\ref{flrew} for this
reason.)
For the distribution of the undegraded light curves of $\gamma$\,Cas the 
turnover is barely discernible. Above 40\,s the interflare distribution
can be fit to an exponential with an e-folding time of 140${\pm 20}$\,s.

\begin{figure}[h!]
\centerline{
\includegraphics[scale=.35,angle=90]
{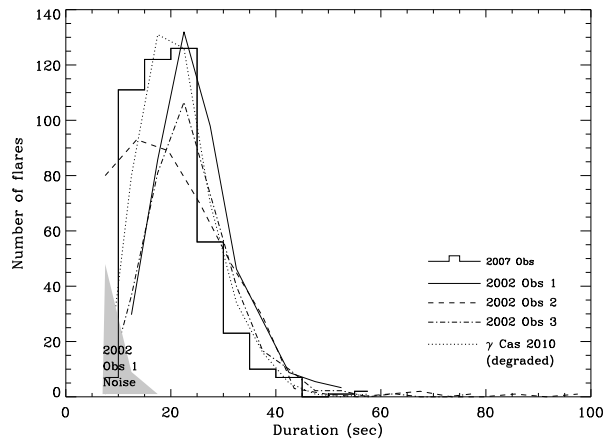}}
\caption{
The distribution of measured flare lifetimes extracted from 
{\it XMM-Newton} light curves of HD\,110342 and (degraded) $\gamma\,$Cas,
taken as a control.
The gray area represents noise fluctuations in 2002 Obs.\,1.
\label{flrlftm}
\vspace*{0.08in}
}
\end{figure}

 The flares in 2007 tend to differ from the 2002-2003 counterparts in their 
distribution of lifetimes.  Taking into account again the noise-generated
false events shown in gray, Figure \ref{flrlftm} shows that the lifetimes of
most flares in 2007 are confined within 10--30\,s. Their distribution is
almost flat within these limits. In contrast, the 2002 (Obs.\,1) and 2003
distributions fall off less rapidly toward short values from a peak at 
20--25\,s.  By comparison, the distribution of
lifetimes of the $\gamma$\,Cas flares (degraded light curve) 
most closely resembled the 2007 data of HD\,110432.
Note also in Fig.\,\ref{flrlftm} that the lifetimes of the HD\,110432 flares 
exhibit an extended, low-amplitude tail out to 80--100\,s. 
In our 2010 $\gamma$\,Cas data the longest-live individual flare was 78\,s.  
However, using the full complement of {\it RXTE/PCU} detectors, 
RS found for $\gamma$\,Cas that flares in 1996 light curves can
occasionally last 150\,s. Although noise can cause two closely spaced
flares to appear as a stronger unresolved one, the reverse is also true.
For example, the strong feature at 26.1 ks in Figure\,\ref{xtelc} 
is actually a 4 minute {\it flare aggregate} followed immediately by a 
weaker group lasting 3 minutes. Given these ambiguities, we do not regard 
the lifetimes and energies of the strongest flares as well defined. 

\vspace*{0.00in}

\subsubsection{Further comments on strong flares from {\it RXTE} data}

  Our review of flares from HD\,110432 was augmented by again degrading
the 2007 {\it XMM-Newton} light curve to match the SNR of the already
low SNR light curve observed by {\it RXTE} in 2010  and again measuring
flares over an equivalent time interval using the {\it flarecount}
program. Using the flare selection criteria chosen in $\S$\ref{flrmeas},
we determined an energy threshold this time of 6 counts.
Our results were, first, that 
the shortest-lived flare events in the degraded time series were 40\,s long. 
Second, we found 61 flares for the 
2007 {\it XMM-Newton} series and 57 flares for the {\it RXTE} series. 
As in the comparison of the higher quality light curves, 
these numbers are indistinguishable from one another. 
Altogether, we conclude that in 2010 HD\,110432's flaring rate was consistent
with its value in 2007. 
As already noted these rates are also consistent with the mean rate 
for $\gamma$\,Cas.

\subsubsection{Color dependence of HD\,110432's flares}
\label{clr}

 In comparing the flares statistics assembled from {\it RXTE/PCA} and 
{\it XMM/EPIC} data, it is relevant to ask whether the respective instrumental 
energy responses bias the statistics. 
We remind the reader that the PCA/PCU's primary energy sensitivity lies in
the range 2--10\,keV whereas the {\it EPIC} detector responses is broader
at both ends. 
Fully 42\% of the 0.6-12\,keV photons collected on HD\,110432 by the
{\it XMM-Newton} detectors lies within the soft 0.6-2\,keV bandpass. 
  We first compared the {\it XMM-Newton} light curve extracted from 2--10\,keV 
({\it RXTE}-like) energies to the 0.6--12\,keV energy range and found 
that basic flare quantities from the two were negligibly different.
This implies that there is little or no color bias in the flare statistics 
between the {\it RXTE} and broader band {\it XMM-Newton} energy ranges.

\begin{figure}
\centerline{
\includegraphics[scale=.35,angle=90]
{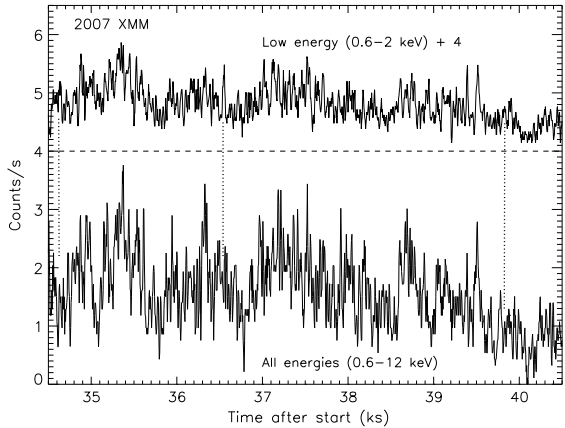}}
\caption{
A stretch in the 2007 light curve of HD\,110432 in two X-ray 
energy bands (plotted in 100\,s bins). Both panels show ubiquitous
rapid fluctuations (flares) and slower variations. Flares usually
occur at all energies, e.g. within the first ks at times
34.62, 34.88, 35.14, 35.37, \& 35.52\,ks.
The vertical 
dotted lines are examples of flare complexes that are stronger 
in the low band than over the full energy range. 
The converse is at least as common.
\label{lowall}
\vspace*{0.08in}
}
\end{figure}

An interesting question is whether light curves in the soft and hard X-ray 
bands always exhibit the same flaring incidence.
Figure\,\ref{lowall} illustrates this question by showing an example
of light curves over several kiloseconds formed from fluxes in the 
soft band (0.6--2\,keV) alone and the full energy range.
L07 noted a correspondence of flares in hard
and soft energy bands. However, there were occasional departures, suggesting 
that flares were somewhat more prevalent in a soft band than a hard one. 
Returning to our flare counting, this time just for 2002 Obs.\,1, 
we found that  $\sim$75\% of the flares identified in the full energy
(0.6--12\,keV) were also recovered in the low energy (0.6--2 keV) one.
In a smaller fraction of instances (15-20\%) the low energy flares were 
stronger, and occasionally significantly so. Three such examples are 
highlighted in Fig.\,\ref{lowall}. Flares could be found only in the soft 
band even less commonly. 
A cross correlation of low (0.6--2\,keV) and high energy
(2--10\,keV) light curves disclosed no time lag between them.
We can add that color differences in general are more common
over timescales of 1\,ks or longer. 
This fact was also noted by
the S12 study of a recent {\it XMM-Newton} light curve of $\gamma$\,Cas.

\section{Other light curve features}

   In compiling our flare statistics, we found that the
distributions of interflare intervals followed apparent
exponential distributions down to the shortest interval between detected
independent flares. Because we know that we miss the unresolved flares 
(``aggregates"), we believe this distribution is consistent with a Poisson 
distribution. This would mean that the flares occur independently of one
another. Put another way, there is no indication of clusters or ``cascades"
of events, such as one finds in true flares on the Sun and late active stars
(e.g., Aschwanden 2011). At times the HD\,110432 light curves
can exhibit as few as two flares per kilosecond, 
for example, during the extended ``drop off" interval discussed next.
Within any one of these comparatively flare-free periods, the basal 
count rate   
can also decrease for several minutes. This behavior is vaguely reminiscent 
of the occurrences of much briefer and cyclical lulls that
often occur for $\gamma$\,Cas (RS, L10). The quasi-periodicity of these  
events is often 3--3.5 or 7 hours (e.g., RS, L10).

\begin{figure}[h!]
\centerline{
\includegraphics[scale=.35,angle=90]
{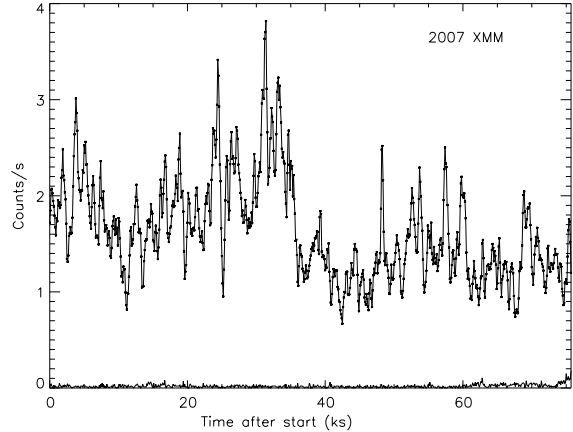}}
\caption{
The 2007 {\it XMM-Newton} light curve for the energy 
range 0.3-12 keV of HD\,110432 and binned to 100 seconds. 
The background count rates are indicated at the bottom.
Notice the sharp decrease at time 35\,ks referred to in the text.
Many positive fluctuations shown are ``flares aggregates" discussed 
in the text. Individual flares generally last for less than 100 seconds 
and are not resolved in this plot. 
\label{lcall}
}
\end{figure}

  Qualitatively new features appearing as rapid and sustained drop offs 
are visible in some of the HD\,110432 light curves, e.g., the end of {\it 
RXTE} Obs.\,2 in Fig.\,\ref{xtelc} and middle of the 2007 {\it XMM-Newton} 
light curve, which is shown in Fig.\,\ref{lcall}. These 
light curves show a step function-like decline (at 35\,ks) to a new low
count rate level. The rates of decline in these events can occur even
over a few minutes, rivaling the decrease and subsequent increase 
rates in the quasi-periodic lulls of the $\gamma$\,Cas light curves.
However, they can last as long as 11 hours, as in Fig.\,\ref{lcall}.  
A similar, though less dramatic drop in counts
can be seen in the {\it RXTE} 
light curve shown in Fig.\,\ref{xtelc}.  We have observed no such 
features in $\gamma$\,Cas light curves.

\section{Discussion and Conclusions}

  The X-ray light curves of 
HD\,110432 exhibit three types of variability: long cycles of a few months,
intermediate variations of a few hours, and rapid flares having lifetimes 
from 1${\frac 12}$ minutes down to several seconds.
These properties are similar to the ubiquitous X-ray variations 
in $\gamma$\,Cas light curves.
The correlation that SRC and SR found between the passages of translucent 
and hot clouds, according to UV continuum and spectral
line time series, 
and increased X-ray fluxes is key evidence for the X-rays being emitted on 
or near the surface of the Be star.  Since the decommissioning in 1997
of the {\it HST Goddard High Resolution Spectrograph,} it has not been 
possible to confirm this implied causality by simultaneous monitoring 
stars of HD\,110432's brightness in both X-ray and UV bands.

  Even so, our {\it RXTE} observations of HD\,110432 over six epochs
in 2010 (Fig.\,\ref{lclong}) suggest the existence of a long
X-ray cycle similar to those of $\gamma$\,Cas.
Combining this feature with the optical light cycle (SB) in 2002, 
we suggest that the long cycles are longer in HD\,110432 than in $\gamma$\,Cas. 
We cannot say yet whether the ranges of the cycles in the two stars can overlap. 

Likewise, the similar statistical properties 
for the long cycle oscillations in HD\,110342 and $\gamma$\,Cas implies a 
constant ratio of 67--80 for the cycle amplitudes for X-ray and optical 
variations.  If this ratio is indeed universal long ``periods" promise to be 
a boon to the optical discovery of new $\gamma$\,Cas analogs. 

 We have also discussed some unique properties of short-term features in
X-ray light curves of HD\,110432. One difference in the HD\,110432 light
is the absence of cyclical, 
several minute-long lulls found in $\gamma$\,Cas light curves.
Rather, at least two of our light curves of HD\,110432
curves exhibit sudden and sustained decreases in X-ray emission.  
The more extreme of these events can be described as drop-offs 
lasting up to several hours. If X-ray flares are the result of
bombardments of high energy particles from an external source onto 
the star, 
the energy reservoir ultimately powering them seems to become
partially depleted for extended intervals. This contrasts
with the recurrent lulls in many of the $\gamma$\,Cas light curves.
This latter description hints of a steady refilling of the reservoir, 
similar to a relaxation cycle.

 In addition, this study and RS have noted differences in the distributions 
of flare energies and lifetimes at various times. One of these is that the
HD\,110432 flares in 2007 tend to be slightly more numerous,
shorter-lived, and more energetic than those in 2002--2003.  
Another peculiarity is the apparent log-linear distribution, 
which {\it seems} to be different from the power law  
relation for solar flares. It is not yet clear how these differences 
can be understood within even a general paradigm.  In any case
we would want to extend the flare count distribution to a larger range
before being satisfied with this conclusion.
At the risk of speculation, we can point out that in the RS picture, 
according to which 
the mean energies of flares is proportional to the input mean beam energy, 
$<$E$>$, the latter energy seem to be slightly higher in 2007. 
If so the beams would penetrate to greater atmospheric densities.
This would account for the more energetic and the 
preponderance of shorter-lived flares observed at that time. 
These differences would require more a complicated explanation in the 
blob accretion picture.


 We have reported that {\it most} flares do not exhibit  color changes 
in HD\,110432, as has also been reported for $\gamma$\,Cas (RS). However, L07
already pointed out the sometime occurrence of soft-energy flares. In this 
study we have found examples of flares that occur at hard energies
but are weak or absent in the soft band. In fewer cases we also discovered
flares more prominent in the soft band.
This also gives us the impression that the energy excitation mechanism 
responsible for flares sometimes operates over a range of energies
although the high end usually dominates. 
In our picture X-ray flares are created by bombardments from without
and are only the manifestation of this high energy input.
RS have discussed how such color changes (including the sometimes observed
inequality $T_{flare}$ $<$ $T_{basal}$) can be produced by introducing 
a small range in a beam energy distribution.

  To sum up, the X-ray variability of HD\,110432 over a variety of timescales
is similar to that of $\gamma$\,Cas.  A comparison of their flare incidences
at different times showed that they can vary by small amounts, but the
degree and timescale of changes from a mean rate are not yet clear.
Likewise, while we are not yet able to corroborate that the optical and 
X-ray long cycles are associated with each other for this star, a 
comparison of their basic properties (perhaps including a common ratio of
the X-ray and optical variations,
L$_{x_{max}}$/L$_{x_{min}}$)/$\triangle V$) for both HD\,110432 and
$\gamma$\,Cas
suggests that the long cycles are common to $\gamma$\,Cas stars at large. 
The cycle lengths likewise vary over time and their mean lengths 
may even differ from star to star. It will be of interest to see how
these lengths are related to general disk characteristics inferred from
optical observations.

\vspace*{0.00in} 

\acknowledgments

The authors wish to thank an anonymous referee, whose comments
contributed greatly to the quality of this paper.
This work was conducted in part from funding of NASA Grant NNX11AF71G 
to Catholic University of America under the Advanced Data Analysis Program.  
RLO acknowledges financial support from the Brazilian agency CNPq ({\it Conselho
Nacional de Desenvolvimento Cient\'ifico e Tecnol\'ogico}) through Universal 
14/2011 Grant 470361/2011-5.

\end{document}